\documentclass[aps,prl,groupedaddress,showpacs,floats,twocolumn]{revtex4}
\usepackage[dvips]{graphicx}
\usepackage{bm}
\begin{document}

\title{Incoherent dynamics of vibrating single-molecule transistors.}

\author{Kevin D. McCarthy}
\author{Nikolay Prokof'ev}
\author{Mark T. Tuominen}

\affiliation{Department of Physics, University of Massachusetts,
Amherst, MA 01003, USA}


\date{\today}

\begin{abstract}
We study the tunneling conductance of nano-scale quantum
``shuttles'' in connection with a recent experiment [H. Park {\it
et al.}, Nature, {\bf 407}, 57 (2000)] in which a vibrating
C$_{60}$ molecule was apparently functioning as the island of a
single electron transistor (SET). While our calculation starts
from the same model of previous work [D. Boese and H. Schoeller,
Europhys. Lett. {\bf 54}, 668 (2001)] we obtain quantitatively
different dynamics.  Calculated I-V curves exhibit most features
present in experimental data with a physically reasonable
parameter set, and point to a strong dependence of the
oscillator's potential on the electrostatics of the island region.
We propose that in a regime where the electric field due to
\textit{the bias voltage itself} affects island position, a
"catastrophic" negative differential conductance (NDC) may be
realized.  This effect is directly attributable to the magnitude
of overlap of final and initial quantum oscillator states, and as
such represents experimental control over  quantum transitions of
the oscillator via the macroscopically controllable bias voltage.
\end{abstract}


\maketitle

\section{Introduction}

Over the past decade, nanotechnology has advanced the ability to
build systems in which chemical self-assembly defines the functional
and structural units of nanoelectronic devices ~
\cite{whitesides,andres,molswitch,sivan}. Since elastic parameters
of organic compounds currently utilized in such work can be much
"softer" than those of semiconductors or metals,  one must
carefully consider the important role \textit{mechanical} degrees
of freedom may play in charge transport devices ~\cite{shekhter}.
It is also fundamentally interesting to consider how the physics
develops as one proceeds from the nanoscale all the way up to
macroscopic electromechanical systems ~\cite{tuominen,erbe}.

A significant impetus in the field of chemically self assembled
nanostructures is in the direction of Coulomb blockade devices, in
which transport is due to the tunneling of a single electron or
Cooper pair through the device via island structures of small
capacitance
 ~\cite{magnus,andres,soldatov,petta}. The quest for systems with
ever smaller capacitance has driven workers to define smaller
structures in such devices, with the result that current
experimental work, such as that modeled here, involves the
utilization of single large molecules as the island structure
through which electronic transport takes place.

The experiment of Ref.~\onlinecite{park} involves the fabrication
of metallic gaps roughly 1 nm in separation
~\cite{electromigration}, and exposure to a solution of $C_{60}$
molecules.  When IV curves were measured, some of the gaps exposed
to $C_{60}$ displayed a robust ~100 meV Coulomb blockade with
charging energy modulation via a third gate terminal (in this case
the gate terminal was the underlying semiconductor substrate).
Also observed was a series of smaller ~5 meV-wide steps
attributable to coupling of the electronic transport through the
island to the quantized vibrations of the center of mass motion of
the molecule in its Van der Waals binding potential nearest to one
of the leads. The steps appear to have some remarkable secondary
features, i.e., asymmetries about zero bias voltage, and unresolved
structure in the intra-step regions of the IV characteristics.

It was suggested in Ref.~\onlinecite{park} that the number of
vibrational steps and their amplitudes are controlled by
Franck-Condon physics. This idea was further developed in
Ref.~\onlinecite{europhys} where Franck-Condon matrix elements
were used to calculate transition rates between different
vibrational states and to solve the master equation for the I-V
curves.   Our work differs from that of Ref.~\onlinecite{europhys}
in the following respects: i.) While we work with the same model
as Ref.~\onlinecite{europhys}, we arrive at a different functional
form for the transition matrix elements, with the result that our
current voltage characteristics disagree quantitatively. ii.)
Because of the unitary nature of the space-shift operators that
must act on the electron tunneling perturbation, we stress the
distinction of shuttle transport into unitary and non-unitary
limits, depending on the degree to which energy conservation
allows the tunneling perturbation can fully connect the Hilbert
spaces of states before and after the island is charged. iii.) We
also argue that a bias voltage dependence of the Franck-Condon
factor should be considered, and that in fact such a dependence
may explain the aforementioned secondary features in the IV
characteristics of Ref.~\onlinecite{park}.  A stronger dependence
is further shown to lead to a ``catastrophic'' negative
differential conductance effect which we will describe in this
paper. The Chalmers group has  modelled the system with classical,
damped oscillations and incoherent electronics \cite{shekhter},
and with classical undamped oscillations and electronic coherence
\cite{fedorets}. Our model is that of incoherent electronics and
quantum mechanical oscillations that are strongly damped.  As
compared with the model of Ref.~\onlinecite{shekhter}, the
molecular island displays its vibrational character not by
shuttling charge at a rate ~$\omega_o$, but in a more
``spectroscopic'' manner, as a sort of fingerprint in the charge
transport physics. Strong dissipation of the oscillator's motion
plays a key role in our incoherent dynamics, in contrast to
 \cite{fedorets}, which assumes \textit{undamped}
classical motion.


\section{Theoretical Model}

\begin{figure}[tbp]
\hspace*{-0.4cm} \includegraphics[width=7cm]{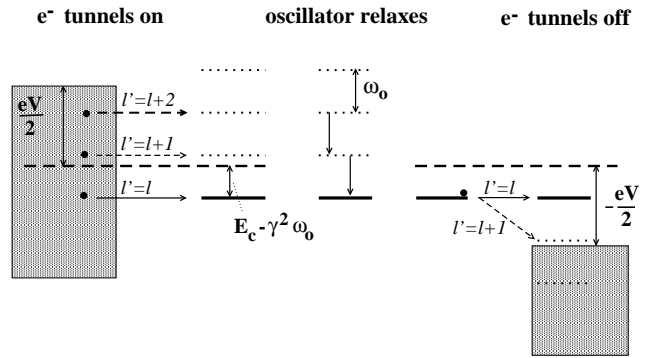}
\vspace*{-4.0cm}
\caption{Schematic of C$_{60}$ transistor system in terms of
combined electronic-vibrational transitions for passage of current
through the device. Dashed arrows indicate transitions in which
quanta of oscillator energy $\hbar \omega_o$ are supplied to the
$C_{60}$ by the tunneling electron}
\label{fig1}
\end{figure}

Fig. ~(\ref{fig1}) shows a depiction of the $C_{60}$ transistor
system.  The physics is assumed to be that of a 3 terminal quantum
dot (including a capacitively coupled gate terminal)  vibrating in
a 1-d quantum harmonic oscillator potential (we disregard
vibrational states in other directions by assuming that their
frequencies are much higher). We write the system Hamiltonian as
follows:
\begin{eqnarray}
\mathit{\hat{H}}=  \mathit{\hat{H}}_{i} + \sum_{s=\pm 1}
\mathit{\hat{H}}_s + \mathit{\hat{H}}_\mathbf{T}
& +&\frac{e^2(\hat{n}-n_o)^2}{2C}-\hat{n} E \hat{x} \nonumber  \\
& + & \hat{V}_\textit{int}^{(\textit{bath})} + \mathit{\hat{H}}_{Bath} \;.
\label{eq:px_hamiltonian}
\end{eqnarray}
The vibrational portion of the island Hamiltonian is that
of a simple harmonic oscillator in
terms of the island center of mass momentum $p$ and coordinate $x$
\begin{equation} \label{eq:island_hamiltonian}
 \mathit{\hat{H}}_{i} = \frac{\hat{p}^{2}}{2m}+\frac{m\omega_o^2\hat{x}^2}{2}
\equiv  \hbar\omega_o(a^{\dag}a+\frac{1}{2}) \;.
\end{equation}
The right ($s=1$) and left ($s=-1$) metallic leads are described
by $\mathit{\hat{H}}_s$ as normal Fermi liquids with constant
Fermi-surface density of states $\rho_F$, and chemical potentials
controlled by the bias voltage. Charge transport between the
island and leads is treated within the tunneling Hamiltonian
approach
\begin{equation}\label{eq:tunnel_hamiltonian}
\mathit{\hat{H}}_\mathbf{T}= t \sum_{s,
k,\sigma}(c^\dag_{sk\sigma}c^{\;}_{i \sigma}
e^{ sx/ \lambda }+h.c.)
\end{equation}
where $c^\dag_{sk\sigma}$ and $c^\dag_{i\sigma}$ are the electron
creation operators in the leads and island respectively. Although
spin is conserved during all tunneling events we need to keep
track of island electronic spin states, since tunneling events in
the two spin channels are correlated. Here $t$ is a constant,
position independent tunneling matrix element, with the island
centered between the metallic leads. Since the island is free to
move, we can also expect an exponential dependence of tunneling on
the center of mass coordinate; the decay of the conduction
electron wave function in vacuum is given by the parameter
$\lambda $.

Next we introduce  the charging (Coulomb) energy of the island
with capacitance $C$ (fourth term) and its dependence on the
island coordinate (fifth term), written in the form of
charge--electric field coupling. The value of $ n_o $ is
controlled by the gate voltage\cite{SCT}. In the present study we
will assume that $C$ is small, and $n_o$ is very close to $1/2$,
so that at all relevant temperatures and bias voltages only two
charge states, $n=0$ and $n=1$, contribute to the dynamics. We
denote the charging energy difference
$E_C=(e^2/2C)[(1-n_o)^2-n_o^2]$. Since the amplitude of zero-point
fluctuations, $u_o = \sqrt{\hbar /2m\omega_o }$, is rather
small for the heavy island (it is estimated as $ \sim 4~pm$ for
the C$_{60}$ molecule \cite{park}) we do not consider terms
non-linear in $x$.

Finally, the last two terms in the Hamiltonian describe an
environment and it's dissipative coupling to the molecule via
$\hat{V}_\textit{int}^{(\textit{bath})}$. Two mechanisms may
contribute here. The vibrational state of the island couples to
the bulk and surface vibrations of leads via standard
displacement-dependent elastic energy. On another hand, island
oscillations produce electrostatic potential fluctuations in the
leads which excite electron-hole pairs. We consider the second
mechanism as dominant because (i) it is Ohmic dissipative while the
phonon coupling is not and vanishes for small energy transfer
\cite{leggett}, and (ii) it is mediated by the unscreened Coulomb
interaction for the charged island. By writing the coupling term as
\begin{equation} \label{dissipation}
\hat{V}_\textit{int}^{(\textit{bath})} = {\hat{x} \over u_o} ~ \sum_{skk'\sigma }
V_{skk'} c^{\dag}_{sk\sigma }c^{\;}_{sk'\sigma}
\end{equation}
we arrive (using golden-rule approximation) at the bath-mediated
transition rates between vibrational states of the island, $|l>$,
\begin{eqnarray}
W_{l \rightarrow \: l+1} &=&
e^{-\hbar \omega_o/k_BT}\: W_{l+1 \rightarrow \: l}  \nonumber \\
                         &=&
K (l+1) {\hbar \omega_o \over e^{\hbar \omega_o/k_BT}-1} \;,
\label{eq:thermal_rate}
\end{eqnarray}
in the form characterized by a single parameter $K \propto (\rho_F
V)^2$. [In a more refined theory one may generalize to consider
$K$ depending on the island charge as well.] We do not see any
obvious reason why this parameter may be anomalously small for
leads with metallic concentration of conduction electrons. Thus
transitions rates between molecular levels of order
$10^{-1}\omega_o$ or $10^{-2} \omega_o $ must be considered as
typical. We performed all our simulations for $K=0.1$.

In what follows we consider the case of weak conductance when
tunneling rates given by $\Gamma^{(0)} = 2 \pi \rho_F t^2 $ are
much smaller than $K \hbar \omega_o $. This condition makes the
theory very simple conceptually because one may ignore
interference between subsequent tunneling events---on time scales
larger than $\hbar /W$ the density matrix of the island reduces to
its diagonal form in the oscillator eigenstates representation and
one may discuss dynamics in terms of probabilities,
$P_{n,\sigma,l}$, of finding a system in a state $|n,\sigma,l>$.
The I-V curve is then obtained from the steady-state solution of
the master equation. This rather standard framework is identical
to the one used in Ref.~\onlinecite{europhys}. All transition
rates for the master equation may be calculated exactly for the
model given above. In what follows we explain how this is done and
analyze various limiting cases by solving equations numerically.
Other work \cite{fedorets} considers the possibility that
electronic coherence persists over timescales larger than
$\sim1/\Gamma^{(o)}$ while assuming \textit{undamped} classical
motion of the oscillator degree of freedom. By contrast, our
master equation approach has strong dissipation of the oscillator
motion as a built-in condition.  In \cite{fedorets} it appears
that the stable limit cycles of oscillation are a result of a
kinetic stability, and not energetic equilibrium of the oscillator
with the environment.

To calculate tunneling matrix elements we have to take care of
final-state interaction effects originating from the $\hat{n} E
\hat{x}$ term, in other words we first have to diagonalize the
Hamiltonian in the absence of tunneling separately for $n=0$ and
$n=1$, and calculate matrix elements between the eigenstates
corresponding to $n$ values differing by $\pm 1$. The resulting
$n=0$ and $n=1$ Hamiltonians are related to one another by a shift
in space of the potential minimum. Technically, it is
convenient to calculate matrix elements in the same basis, and
account for the difference between initial and final Hamiltonians
by constructing a unitary transformation relating the
corresponding eigenfunctions. This is the essence of the
Franck-Condon principle. In our case we have to deal with the
shifted-oscillator basis set which is a textbook problem.
Oscillator states with $n=1$ and $n=0$ are related by
\begin{equation} \label{unitary}
|n=1,l> = e^{\hat{S}}\: |n=0,l>\;,
\end{equation}
where
\begin{equation}\label{eq:soperator}
\hat{S}\equiv \gamma \: (a^{\dag}-a) \;,
\end{equation}
with dimensionless coupling parameter
\begin{equation}\label{gamma}
\gamma=\frac{eE\textit{u}_o}{\hbar\omega_o}\;.
\end{equation}
The ground state energy levels are shifted by $-\gamma^2 \hbar
\omega_o$.

Since the oscillator Hamiltonian suddenly changes during a
tunneling event, the above unitary transformation is all we need
in order to calculate matrix elements because it projects $n=1$
states into the $n=0$ Hilbert Space. An effective form of the
tunneling Hamiltonian (in the $n=0$ representation) is given now
by
\begin{equation}\label{eq:dressed tunneling hamiltonian}
\widetilde{H}_T= t \sum_{sk\sigma} ( e^{s\alpha
(a^{\dag}+a)}\:e^{\hat{S}}\: c^{\dag}_{sk\sigma}
\:c^{\;}_{\textit{i}\sigma} + h.c.)
\end{equation}
where dimensionless $\alpha = u_o/\lambda $ measures the ratio
between the molecule zero-point displacement and electron
localization length. The energy shift term $-\gamma^2 \:\hbar
\omega_o$ must be taken into account when writing the energy
conservation law for all transitions. Specifically, when an island
charges/discharges, the tunneling electron will gain/lose an
energy equal to the shift. The above considerations are readily
generalized for any relevant values of charge states.

So the picture we have is that of the combined
electronic-vibrational system with the tunneling Hamiltonian
giving dynamics to both degrees of freedom. In the
realistic limit of $ W \gg \Gamma^{(0)} $, we can assume
incoherent dynamics and move to a master equation formulation of
the the occupation probabilities of quantum states of the
transistor:
\begin{eqnarray}
\frac{dP_{n,\sigma,\mathit{l}}}{dt}=
&-&
P_{n,\sigma,\mathit{l}} \sum_{n',\sigma',\mathit{l}'}
\Gamma_{n,\sigma,\mathit{l} \rightarrow \: n',\sigma',\mathit{l}'} \nonumber \\
&+&
\sum_{n',\sigma',\mathit{l}'} P_{n',\sigma',\mathit{l}'}
\Gamma_{n',\sigma',\mathit{l}' \rightarrow n,\sigma,\mathit{l}}
\;.
\label{master}
\end{eqnarray}
By considering transitions through only one side of the island
(e.g., the right hand side) we obtain the current as
\begin{equation}\label{eq:current}
I= e \sum_{\sigma,\mathit{l},\mathit{l}'} \left[
P_{1,\sigma,\mathit{l}} \Gamma^{(R)}_{1,\sigma,\mathit{l}
\rightarrow 0,\mathit{l}'}
-P_{0,\mathit{l}}(\Gamma^{(R)}_{0,\mathit{l} \rightarrow
1,\sigma,\mathit{l}'} \right] \;.
\end{equation}

We have two kinds of transitions in Eq.~(\ref{master}), the thermal rates from
(\ref{eq:thermal_rate}), which change only the vibrational state ($n'=n$),
and the charge-transfer rates, $\Gamma = \Gamma^{(L)}+\Gamma^{(R)}$, with
\begin{eqnarray}
 \Gamma_{ n,\sigma,\mathit{l}  \rightarrow n',\sigma',\mathit{l}'}
= {2\pi \over \hbar }\sum_{fi} \rho_{i}^{(eq)}
\bigg| \langle f,n',\sigma',\mathit{l}' \big| \mathit{\widetilde{H}}_{T}
\big| i,n,\sigma, \mathit{l} \rangle \bigg|^2 & & \nonumber \\
\times \delta(E_f-E_i+\epsilon_{n',\sigma',\mathit{l}'}-\epsilon_{n,\sigma,\mathit{l}})
\;;~~~(n'\ne n)\;,\;\;\;\;\;& &
\end{eqnarray}
where $n,\sigma$ and  $n', \sigma'$ are the initial and final
electronic state indices for the island, $\mathit{l}, \mathit{l}'$ are likewise the initial and
final vibrational state indices, and $i,f$ stand for the initial and final
states of the leads which are considered to be in the state of
thermal equilibrium ($\rho_{i}^{(eq)}$ is the equilibrium density matrix of the leads).
We also assume the leads to be symmetrically voltage biased,
i.e. $+V/2$ on the left lead and $-V/2$ on the right one, to avoid unnecessary
complications. For the same reason
we do not consider the case when tunneling rates
$\Gamma^{(0)}$ are different for the left and right leads.
Explicit expressions for charge-transfer rates are:
\begin{eqnarray}
\Gamma_{0,0,\mathit{l} \rightarrow 1,\sigma,\mathit{l}'}^{(s)} &=&
\Gamma^{(0)}\: \left| \bm{O}^{(s)}_{\mathit{l}' \mathit{l}}
\right|^{2}\: f(\delta E)  \\*
\Gamma_{1,\sigma, \mathit{l}' \rightarrow 0,0,\mathit{l} }^{(s)} &=&
\Gamma^{(0)}\: \left|
\bm{O}^{(s)}_{\mathit{l} \mathit{l}'} \right|^{2} \: \left[ 1- f(\delta E) \right] \\
\delta E &=& E_c+\frac{seV}{2}+\hbar \omega_{o}(\mathit{l}' -\mathit{l}-\gamma^2)  \;,
\end{eqnarray}
(where $f(x)$ is the Fermi distribution function)
with oscillator matrix elements
\begin{eqnarray}
\bm{O}^{(s)}_{\mathit{l}' \mathit{l}} & =&
\bm{O}^{(s)}_{\mathit{l}\mathit{l}'}=
e^{-s\alpha\gamma +\frac{1}{2}(\alpha^2-\gamma^2)} \sqrt{\mathit{l}'!\mathit{l}!} \nonumber \\
&\times &
\sum_{m=0}^{min(\mathit{l},\mathit{l}')}
{ \left( s\alpha+\gamma \right)^{\mathit{l}'-m}
 \left( s\alpha-\gamma \right)^{\mathit{l}-m} \over
 (\mathit{l}-m)!   (\mathit{l}'-m)!     m!        } \;.
\label{matrix_elements}
\end{eqnarray}
For all simulations we measure energies and voltages in units of
$1~meV$, and current in units of $e \Gamma^{(0)}$.

In calculating IV characteristics from Eqns. (\ref{master}) and
(\ref{eq:current}), we have to adopt a model for the bias voltage
dependence of the electric field $E$ at the position of the
island, see (\ref{eq:px_hamiltonian}). Phenomenologically, one
might assume there is bias voltage independent component, $E_0$,
which includes effects of the gate, charged impurities, image
charges and screening, device geometry, etc., and bias dependent
part proportional to $V/d$ where $d$ is the separation between the
leads. Since the electric field for zero and non-zero V is
calculated for the same charge and position of the island, we do
not see any physical reason why the dependence of the field on $V$
might be non-analytic. While this effect was implicit in the
classical oscillator of \cite{shekhter}, it was not considered in
~ \cite{park,europhys}. We argue below that it is precisely this
 term that may help to explain some of the
anomalous features in the data of \cite{park}, and that negative
differential conductance (NDC) may arise in systems in which this
terms dominates over $E_0$.  Indeed, this bias voltage dependence,
along with gate voltage dependence of $E_0$, represent macroscopic
"knobs" with which to control the quantum mechanical transition
matrix element between initial and final oscillator states. Thus
we have to assume that coupling parameter $\gamma $ is a function
of $V$, and write it as
\begin{equation} \label{gammaV}
\gamma\equiv\frac{eEu_o}{\hbar \omega_o}
\equiv c_1+(eV/\hbar \omega_o)\: c_2
\end{equation}
where $c_1$ and $c_2$ parameterize the relative contributions of
bias voltage independent and bias voltage dependent electric
fields.

\begin{figure}[tbp]
\includegraphics[width=6.5cm]{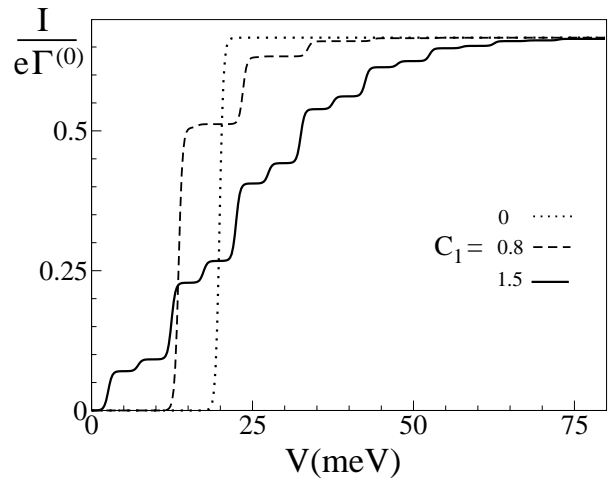}
\vspace*{-1.8cm}
\caption{ IV characteristics for $c_2=0, c_1=0.0,~
 0.8,~ 1.5$ with the voltage offset for each curve defined by
the parameter $E_c= 10~meV$ (see text) }
\label{fig2}
\end{figure}

\vspace*{0.2cm}
\begin{figure}[tbp]
\includegraphics[width=6.5cm]{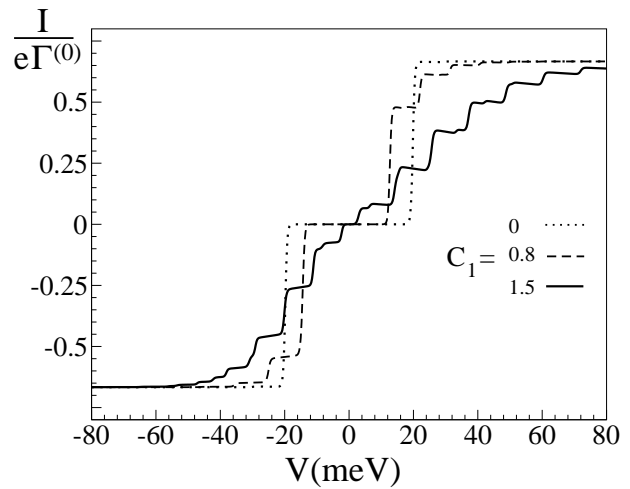}
\vspace*{-1.8cm} \caption{ IV characteristics for finite
$c_2=0.005$, $c_1=0,~0.8,~1.5$, and $E_c= 10~meV$.}
\label{fig3}
\end{figure}

\section{Simulation Results}


Since for heavy islands we expect $\alpha \ll 1$ to be a typical
situation, we start the analysis from the $\alpha=0 $ case first.
When voltage dependence of $\gamma $ is ignored (as in
Ref.~\onlinecite{europhys}), i.e. $c_2=0$, we expect that in the
limit $V/\hbar \omega_o >> \gamma^2$ all I-V curves must saturate
to the ideal value $ I_0=2e\Gamma ^{(0)}/3$. The reason for that is
in the unitary nature of Franck-Condon factors; for zero $\alpha$
we have $\bm{O}^{(s)}_{\mathit{l}' \mathit{l}} = <\mathit{l}'| e^S
| \mathit{l}>$, and thus when all excitation channels are
contributing to the current, we have $\sum_{l'} |
\bm{O}^{(s)}_{\mathit{l}' \mathit{l}}|^2 \equiv 1$.
Fig.~(\ref{fig2}) shows calculated IV characteristics in the
strong dissipation limit for $c_2=0, c_1=0.0,~0.3,~0.8,~1.5$. We
have chosen (for all plots) $\hbar\omega_o=5~meV$,
$k_BT=0.15~meV$, to correspond to the experimental data.

There are several observations to be made. First, steps appear at
roughly $10~meV$ intervals, corresponding to $eV/2=k
\hbar\omega_o$, with $k=0,1,2,...$. Note that it is only energy
conservation in tandem with non-zero Franck-Condon transition matrix
elements that gives multiple steps. Second, the current through the device
saturates to the same limit for all values of $c_1$.
Also, since the action of the unitary operator is to project the initial
oscillator state (in the strong dissipation limit, this initial
state is always the ground state of the oscillator) into the
eigenbasis of a space-shifted Hamiltonian, we see that the larger
the shift parameter $\gamma$, the more access the system has to
transitions into states above the ground state of the shifted
basis, resulting in more visible steps in the IV characteristic.
This is in contrast to the results of Ref.~\onlinecite{europhys}, Fig.~3,
where many steps are present when $c_1=0$ and the IV curves do
not satisfy the unitary limit requirement. The origin of discrepancy
is hard to identify since the transition matrix elements were not
given explicitly in  Ref.~\onlinecite{europhys}.

We note, that for $c_1>1$ the structure of the curve depends on the
value of the $2(E_c -\gamma^2 \hbar \omega_o)/\hbar \omega_o$
parameter because different number of vibrational states are
accessible for the electron to enter and to leave the island.
These states appear at $2 \hbar \omega $ intervals and form two
separate sets which do not coincide if  $2(E_c/\hbar \omega_o
-\gamma^2)$ is not integer. When one of the sets reaches its
unitary limit before the other set starts conducting (as in the
cases $c1=0.0, 0.3$  in Fig.~(\ref{fig2})) we observe only the
second set (thus curves with large offset show only one set of
steps). For small values of $c_1$ conductance is dominated by just
one or two states and the second set has no visible effect on the
curve (``fades away'').  For the largest value of $c_1=1.5$ we see
that neither set of steps (entering or leaving island) reaches a
unitary limit before the other and hence both sets of steps are
resolved.

\begin{figure}[tbp]
\includegraphics[width=6.5cm]{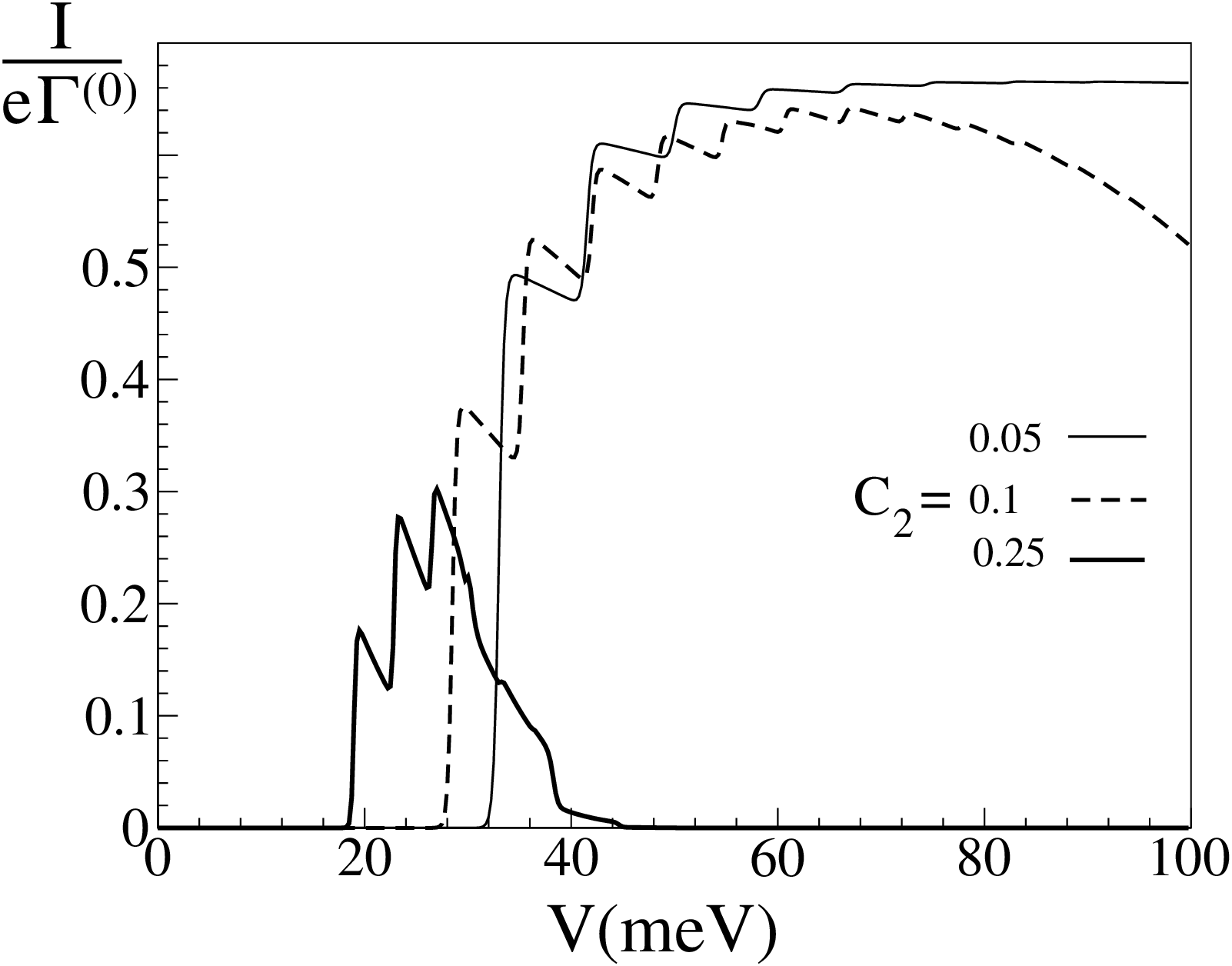}
\vspace*{-1.8cm} \caption{ IV characteristics for $c_1=0.5$,
$c2=0.05,~0.1,~0.25$, and $E_c= 20~meV$}
\label{fig4}
\end{figure}

When voltage dependence is accounted for in $\gamma $,
Eq.~(\ref{gammaV}), the I-V curves become asymmetric. The obvious
effect of non-zero $c_2$ is that now the $\textit{bias voltage}$
affects the unitary transformation. In particular, we note
negative differential conductance in the intra-step regions on the
positive side of bias, while the intra-step regions at negative
bias have positive differential conductance, see
Fig.~(\ref{fig3}). This can be explained in the following way: in
the intra-step region on the positive side of bias, as voltage
increases, the shift parameter $\gamma$ increases, reducing the
overlap between the initial oscillator state and the final,
shifted state (or states), and hence \textit{quantum mechanically}
constricting that particular $l\to l'$ channel.  For negative bias
voltages, we note that the constant electric field parametrized by
$c_1$ and the voltage dependent electric field parametrized by
$c_2$ \textit{work against one another}, such that orthogonality
between initial and final oscillator states is overall reduced.
These I-V characteristics display what appears to be salient in
the data of \cite{park}, namely an asymmetry about zero voltage in
differential conductance, particularly in the intra step regions.

\begin{figure}[tbp]
\includegraphics[width=6.5cm]{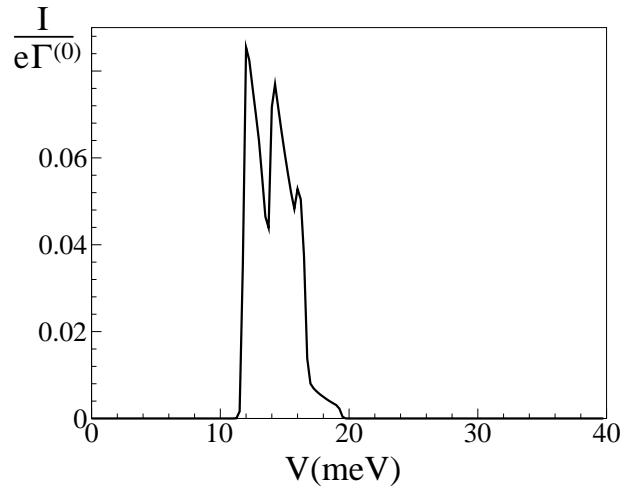}
\vspace*{-1.8cm} \caption{ Conductance resonance due to
"catastrophic" negative differential conductance effect for
$c_1=0.5$, $c2=0.5$, and  $E_c=20~meV$}
\label{fig5}
\end{figure}

Fig.~(\ref{fig4}) shows I-V curves for
$c_2=0.05,~0.1,~0.25$, with $c_1=0.5$ and $\alpha =0$ with
charging energy $E_c=20 meV$. Here we see what we refer to as
"catastrophic" negative differential conductance. Although the
tunneling operator is still unitary, as $c_2$ increases, the
energy-shift of the final oscillator basis grows strongly (as
$\sim -\gamma^2 \hbar \omega_o$), with the result that the initial
state has its largest projection onto states high in the
final-state spectrum. As the bias voltage increases, the
orthogonality effect is more and more important, until most of the
(oscillator) overlap is in a region where energy cannot be
conserved, and so no channels for oscillator transitions remain,
giving nearly zero current through the island. Finally,  when
$|E_c-(c_1+(V/\hbar \omega_o)~c_2)^2 \hbar \omega_o | $ gets
larger than $eV/2$ the number of states available for the electron
to tunnel off the island is reduced to zero. Fig.~(\ref{fig5})
shows how this effect can produce a sharp conductance resonance.

\begin{figure}[tbp]
\includegraphics[width=6.5cm]{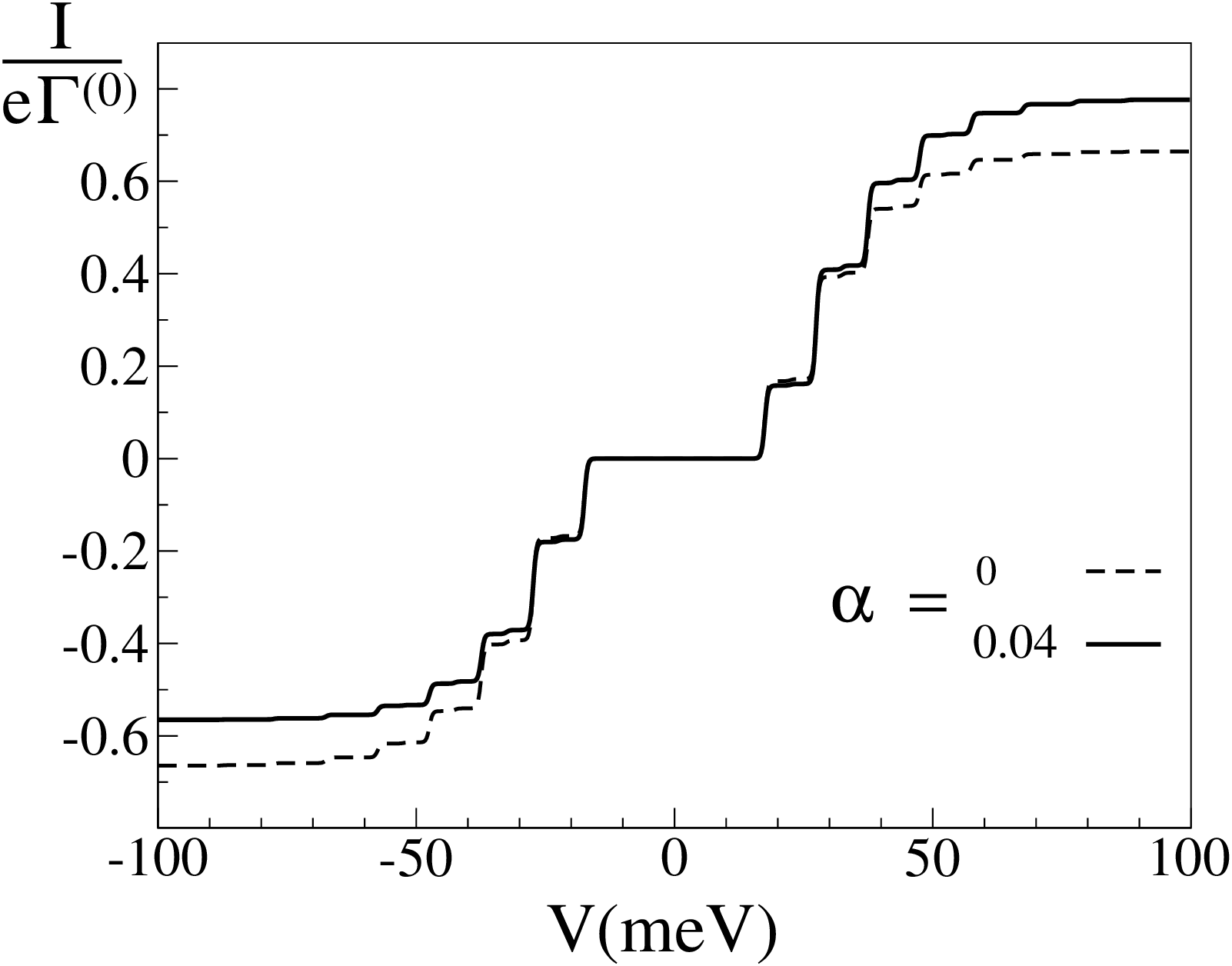}
\vspace*{-1.8cm} \caption{ IV characteristics with $c_2=0$;
$c_1=1.5$ and $\alpha=0,~0.04$, and $E_c=20~meV$ }
\label{fig6}
\end{figure}

Fig.~(\ref{fig6}) shows the effect of a non-zero $\alpha$. We have
set $\alpha=0$, $c1=1.5$, and $c_2=0$ for one of the plots, and
$\alpha=0.04$, $c1=1.5$, and $c_2=0$ for the other. $\alpha=0.04$
is a physically reasonable value which for the C$_{60}$ case
corresponds to the localization length $\lambda \sim 1~A$. Curves
with finite $\alpha$ differ only slightly from the $\alpha \to 0$
case in the magnitude of current at various steps and asymmetry
around $V=0$.  This is indicative of the manner in which $\alpha$
destroys the unitarity condition $\sum_{l'} |
\bm{O}^{(s)}_{\mathit{l}' \mathit{l}}|^2 \equiv 1$. Since we
assume here that zero-charge island states are not shifted, the
system conducts better when charged states are shifted towards
lower voltage lead. We note that charge transfer physics discussed
here radically differs from the semiclassical limit of
large-amplitude vibrations presented in
Refs.~\onlinecite{shekhter} ,~\onlinecite{fedorets} where electron
tunneling was considered to be the fastest process at least for
some values of $x$. In our case, finite $\alpha $ does not
necessarily mean larger conductance at all voltages. Also, in the
semiclassical limit of large-amplitude vibrations the answer
crucially depends on dissipation---the vibration amplitude is set
by the balance between the energy gained from charge-transfer
processes and energy lost to the environment (in the limit of
large $c_1,c_2$ and relatively large $\alpha $).

\section{Conclusion}
Even at the level of the Hamiltonian one can easily see that the
dynamics are controlled by many parameters. Although the master
equation formally solves the problem for an arbitrary parameter
set (as long as dissipation is strong enough), it is difficult to
cover all possibilites single paper, even with some of the
simplifying assumptions used above.  That said, we compare with
the $C^{60}$ experiment and find that the IV characteristics can
be more or less captured by the $c_1\sim 1, c_2\sim 0.05$ ,
$\alpha\sim 0.1$ parameter set. It does appear in \cite{park} that
there may be features (NDC in the intra-step regions) that reflect
the physics of the  $c_2\neq 0$ case (see Fig.~(\ref{fig3})) when
the space-shift in the oscillatory potential immediately after a
tunneling event depends upon the bias voltage. It is intriguing to
note that with $c_2\ne 0$ in our model, we have a situation in
which the bias voltage, a macroscopically controllable variable,
can have an observable influence on the quantum matrix elements
which govern the dynamics of the interacting, quantum mechanical
degrees of freedom for the system. Future experiments on
molecular-based transistors may further explore the existence of
such effects, and point the way toward greater control over
quantum transitions. The electronically coherent model
\cite{fedorets}, also seems able to reproduce primary features in
the experimental IV curves, but come from otherwise qualitatively
different behavior (an classically oscillating island with
undamped, stable amplitude $\sim 1$ Angstrom, compared with our
strongly damped, near zero point motion ($\sim $picometers)). We
find that the most interesting physics in our model is the
(possibly catastrophic) negative differential conductance effect
arising from the experimentally controlled Franck-Condon
transition matrix elements.
\section{Acknowledgment}
The authors would like to thank R.V. Krotkov for numerous
discussions of shuttling physics.

This work was supported by the National Science Foundation
under Grants DMR-0071767 and DMR-0071756.

\end{document}